\documentclass[11pt, a4paper]{article}
\usepackage[top=0.8in, bottom=0.8in, left=1.0in, right=1.0in]{geometry}
\usepackage{booktabs}
\usepackage{multirow}
\usepackage{setspace}
\usepackage{natbib}
\usepackage{authblk}
\usepackage[pdftex]{graphicx}
\usepackage{amsmath}
\usepackage{amsthm}
\usepackage{mathtools}
\usepackage{amsfonts}
\usepackage{ascmac}
\usepackage{amssymb}
\usepackage{bm}
\usepackage{color}
\usepackage{enumitem}
\usepackage{mathrsfs}
\usepackage[dvipdfmx,
	colorlinks = true, 
	anchorcolor=black,
	linkcolor = blue,
	citecolor = blue, 
	filecolor  = black, 
	urlcolor   = blue
]{hyperref}
\usepackage{comment}
\usepackage{algorithmic, algorithm}

\usepackage{breqn}
\allowdisplaybreaks[1]

\newcommand{\biblist}{\begin{list}{}
{\listparindent 0.0cm \leftmargin 0.50cm \itemindent -0.50 cm
\labelwidth 0 cm \labelsep 0.50 cm
\usecounter{list}}\clubpenalty4000\widowpenalty4000}
\newcommand{\ebiblist}{\end{list}}

 \theoremstyle{plain}
\newtheorem{thm}{Theorem}

\theoremstyle{definition}

\theoremstyle{remark}
\newtheorem*{rem}{Remark}

\newcommand{\tr}{{\rm tr}}

\newcommand{\diag}{{\rm diag}}

\newcommand{\pd}{\partial}

\newcommand{\MSE}{{\rm MSE}}
\newcommand{\E}{\mathbb{E}}

\newcommand{\argmin}{\mathop{\rm arg~min}\limits}

\title{\bf Improving the accuracy of estimating indexes \\in contingency tables using Bayesian estimators}
\author[1]{Tomotaka Momozaki}
\author[1]{Koji Cho}
\author[3]{Tomoyuki Nakagawa}
\author[2,3,4]{Sadao Tomizawa}

\affil[1]{Department of Information Sciences, Graduate School of Science and Technology, Tokyo University of Science}
\affil[2]{Department of Information Sciences, Faculty of Science and Technology, Tokyo University of Science}
\affil[3]{School of Data Science, Meisei University}
\affil[4]{Department of Information Science, Meisei University}
\date{Last update: \today}

\begin{document}
\maketitle
%=abstract=========================================================
\begin{abstract}
In contingency table analysis, one is interested in testing whether a model of interest (e.g., the independent or symmetry model) holds using goodness-of-fit tests.
When the null hypothesis where the model is true is rejected, the interest turns to the degree to which the probability structure of the contingency table deviates from the model.
Many indexes have been studied to measure the degree of the departure, such as the Yule coefficient and Cram{\'e}r coefficient for the independence model, and Tomizawa's symmetry index for the symmetry model.
The inference of these indexes is performed using sample proportions, which are estimates of cell probabilities, but it is well-known that the bias and mean square error (MSE) values become large without a sufficient number of samples.
To address the problem, this study proposes a new estimator for indexes using Bayesian estimators of cell probabilities.
Assuming the Dirichlet distribution for the prior of cell probabilities, we asymptotically evaluate the value of MSE when plugging the posterior means of cell probabilities into the index, and propose an estimator of the index using the Dirichlet hyperparameter that minimizes the value.
Numerical experiments show that when the number of samples per cell is small, the proposed method has smaller values of bias and MSE than other methods of correcting estimation accuracy.
We also show that the values of bias and MSE are smaller than those obtained by using the uniform and Jeffreys priors.
\end{abstract}


\medskip

%=introduction=========================================================
\section{Introduction}
\label{sec:intro}
For two-way contingency tables, an analysis is generally performed to determine whether the independence between the row and column classifications holds. 
Meanwhile, for the analysis of square contingency tables with the same row and column classifications, there are many issues related to symmetry rather than independence. 
This is because, in square contingency tables, there is a strong association between the row and column classifications.
\cite{bowker1948test} proposed the symmetry model.
Many other models for symmetry and asymmetry have been proposed, such as marginal homogeneity (\citealp{10.1093/biomet/42.3-4.412}), quasi-symmetry (\citealp{AFST_1965_4_29__77_0}), conditional symmetry (\citealp{10.1093/biomet/65.2.413}), and diagonals-parameter symmetry (\citealp{goodman1979multiplicative}).
For details, see \cite{tahata2014symmetry}.

In the analysis of two-way contingency tables, the degree of departure from independence is measured using indexes between the row and column variables.
These indexes include Yule's coefficients of association and colligation (\citealp{yule1900vii, yule1912methods}), Cram\'{e}r's coefficient (\citealp{cramir1946mathematical}), and Goodman and Kruskal's coefficient (\citealp{goodman1954measures}).
For details, see \cite{bishop2007discrete} and \cite{agresti2013categorical}.
\cite{tomizawa1997generalized} generalized Goodman and Kruskal's coefficient via the power-divergence (\citealp{cressie1984multinomial}).
\cite{tomizawa2004generalization} also generalized Cram\'{e}r's coefficient via diversity index.

In addition, in the analysis of square contingency tables with the same row and column classifications, we are interested in measuring the degree of departure from symmetry or asymmetry.
Over the past few years, many studies have proposed indexes to represent the degree of departure from symmetry or asymmetry.
For square contingency tables with nominal categories, \cite{tomizawa1998power} and \cite{tomizawa2001generalized} proposed indexes based on the power-divergence and diversity index (\citealp{patil1982diversity}) to represent the degree of departure from the symmetry and marginal homogeneity models, respectively.
For square contingency tables with ordered categories, \cite{tomizawa2001theory} and \cite{tomizawa2005power} proposed indexes based on the power-divergence and diversity index to represent the degree of departure from the symmetry and diagonals-parameter symmetry models, respectively.

Although these indexes are estimated using sample proportions, which are typical estimators of cell probabilities, it is well known that the bias and mean squared error (MSE) values become large when the number of samples is not sufficient relative to the number of cells.
To solve this problem, \cite{tomizawa2007improved}, \cite{tahata2008improved}, and \cite{tahata2014refined} derived higher orders of bias and performed bias corrections.
They showed numerical experiments that the improved estimators approached the true values of the indexes faster than the estimators with sample proportions as the sample size increases.
However, although the improved estimators certainly reduce the value of bias, they do not necessarily reduce the value of MSE, that is, the variances of the estimators may become large.
In addition, due to the bias correction term, the range of possible values for these improved estimators is not equal to that of their corresponding indexes.
For example, the value of \cite{tomizawa1998power}'s symmetry index lies between 0 and 1, but the value range of the improved estimator is beyond the range of 0 to 1.
If the value of the improved estimator is outside the value range of the index, it would be difficult for analysts to interpret the value.
It is also difficult to derive the asymptotic distribution of the improved estimator.
Therefore, it is noted that the uncertainty quantification of the index based on the improved estimator may not be possible.

To solve further problems of the improved estimator in the estimation of indexes, we propose a newly inference method of indexes that improves the problems of their improved estimator, based on their idea of bias correction by deriving higher order of bias in the index itself rather than in the cell probabilities.
Namely, we asymptotically evaluate the MSE of the estimator of index using the posterior means of the cell probabilities with the Dirichlet prior instead of sample proportions, and derive the Dirichlet hyperparameter that minimizes the MSE.
Our proposed estimator of indexes is constructed based on the posterior means of the cell probabilities with the derived Dirichlet parameter.
The uncertainty quantification of the indexes in our proposed method can be easily performed by using the Monte Carlo simulation.

In a sense, this study may solve the problem of which the Dirichlet parameter can be used for precise estimation of indexes.
There are many studies on the choice of Dirichlet parameters for estimating cell probabilities.
One of the most famous is the uniform prior Dirichlet($1,\ldots,1$), which originated in \cite{bayes1763lii}, and the Jeffreys prior Dirichlet($1/2,\ldots,1/2$), derived from the invariance rule by \cite{jeffreys1946invariant}.
With $k$ the number of cells, Dirichlet($1/k, \ldots, 1/k$) was originally suggested by \cite{perks1947some} and recommended as an ``overall objective'' prior by \cite{berger2015overall}.
\cite{fienberg1972choice} evaluated the variation of the risks of the posterior means of the cell probabilities with respect to the Dirichlet parameters.
\cite{fienberg1973simultaneous} derived the Dirichlet parameters that asymptotically minimize the MSEs of the posterior means of the cell probabilities.
Other studies, such as \cite{tuyl2018method}, have discussed the choice of Dirichlet parameters in various situations, such as when there are many zero cells.
Thus, there are many studies on the method of selecting the Dirichlet parameters in estimating cell probabilities other than those mentioned above, but to our knowledge, there is no study that discusses how to select the Dirichlet parameters to improve the accuracy of the estimation of indexes in contingency tables.
Our contribution is not only to improve the accuracy of the estimation of index in contingency tables, but also to provide a method for selecting the Dirichlet hyperparameter when using Bayesian estimators of cell probabilities for the estimation of index.

This paper is organized as follows.
Section \ref{sec:MSE} asymptotically evaluates the MSE of the estimator of the index with the posterior means of the cell probabilities and derives the Dirichlet parameter that asymptotically minimizes the MSE.
Section \ref{sec:experiments} shows that the proposed estimators can reduce the bias and MSE more than other estimators in the numerical experiments.
Section \ref{sec:concluding} presents the concluding remarks. 

%=sec:MSE=========================================================
\section{Dirichlet Parameter that Asymptotically Minimizes the MSE}
\label{sec:MSE}
Consider an $r \times c$ contingency table.
Suppose that $\bm{n} = (n_{11}, n_{12}, \ldots, n_{1c}, n_{21}, \ldots, n_{rc})^\top$ is a random vector with a multinomial distribution:
\begin{equation*}
p(\bm{n} \mid \bm{p}) = \frac{n!}{\prod_{i,j} n_{ij}!} \prod_{i,j} p_{ij}^{n_{ij}},
\end{equation*}
where $n = \sum_{i,j} n_{ij}$, $p_{ij}$ is the probability that an observation falls in the $i$th row and $j$th column of the table ($i=1,\ldots,r; j=1,\ldots,c$), $\bm{p} = (p_{11}, p_{12}, \ldots, p_{1c}, p_{21}, \ldots, p_{rc})^\top$, and $\bm{b}^\top$ is the transpose of $\bm{b}$.
Let $\bm{p}$ have a Dirichlet prior density
\begin{equation*}
p(\bm{p} \mid \alpha) = \frac{\Gamma(rc\alpha)}{(\Gamma(\alpha))^{rc}} \prod_{i,j} p_{ij}^{\alpha-1},
\end{equation*}
where $\Gamma(\cdot)$ is the Gamma function.
In this case, the posterior distribution of $\bm{p}$ is
\begin{equation*}
p(\bm{p} \mid \bm{n}) = \frac{\Gamma(\sum_{i,j}(\alpha+n_{ij}))}{\prod_{i,j} \Gamma(\alpha+n_{ij})} \prod_{i,j} p_{ij}^{n_{ij}+\alpha-1},
\end{equation*}
so the posterior mean of $\bm{p}$ is
\begin{equation*}
\hat{\bm{p}}^{(\alpha)} = (\hat{p}_{11}^{(\alpha)}, \hat{p}_{12}^{(\alpha)}, \ldots, \hat{p}_{1c}^{(\alpha)}, \hat{p}_{21}^{(\alpha)}, \ldots, \hat{p}_{rc}^{(\alpha)})^\top,
\end{equation*}
where
\begin{equation*}
\hat{p}_{ij}^{(\alpha)} = \frac{n_{ij} + \alpha}{n + rc\alpha}.
\end{equation*}
When $\alpha = 0$, the posterior mean $\hat{p}_{ij}^{(\alpha)}$ corresponds to the sample proportion $\hat{p}_{ij} = n_{ij} / n$.
When we adopt the squared distance from the estimator to $\bm{p}$ as the loss function, $\hat{\bm{p}}^{(\alpha)}$ is the Bayes estimator of $\bm{p}$.

Let a function $f(\cdot)$ denote an index in contingency tables.
Many indexes in contingency tables that have been proposed thus far are defined as functions of $\bm{p}$.
Therefore, the value of the index $f(\bm{p})$ is estimated using $f(\hat{\bm{p}})$, where $\bm{p}$ is replaced by the sample proportions $\hat{\bm{p}}$ in $f(\bm{p})$.
In this study, instead of $f(\hat{\bm{p}})$, $f(\hat{\bm{p}}^{(\alpha)})$, where $\bm{p}$ is replaced by $\hat{\bm{p}}^{(\alpha)}$ in $f(\bm{p})$ is considered as an estimator of an index, and in order to improve the accuracy for estimating the index, we asymptotically evaluate the MSE of $f(\hat{\bm{p}}^{(\alpha)})$ and derive the Dirichlet parameter that minimizes it.

First, we consider the Dirichlet parameter as follows.
\begin{equation*}
\alpha^* = \argmin_{\alpha} \lim_{n\to\infty} n^2 \MSE[f(\hat{\bm{p}}^{(\alpha)})].
\end{equation*}
Here, the following theorems hold.
\begin{thm}
\label{thm:mse}
Suppose that $f(\cdot)$ is at least four times differentiable at $\bm{p}$.
The MSE of $f(\hat{\bm{p}}^{(\alpha)})$ is expressed as
\begin{align*}
\MSE[f(\hat{\bm{p}}^{(\alpha)})] 
=& \frac{1}{n^2} \left( A_1 \alpha^2  - 2 A_2 \alpha \right) + (\mbox{terms independent of $\alpha$}) + o(n^{-2}),
\end{align*}
where
\begin{align*}
A_1 =& \tr\left[ \left( \frac{\pd f(\bm{p})}{\pd \bm{p}} \right) \left( \frac{\pd f(\bm{p})}{\pd \bm{p}^\top} \right) (rc\bm{p} - \bm{1}_{rc}) (rc\bm{p} - \bm{1}_{rc})^\top \right], \\
A_2
=& \frac{1}{2} (rc\bm{p} - \bm{1}_{rc})^\top \left( \frac{\pd f(\bm{p})}{\pd \bm{p}} \right) \tr\left[ \left( \frac{\pd^2 f(\bm{p})}{\pd \bm{p} \pd \bm{p}^\top} \right) (\diag(\bm{p}) - \bm{p} \bm{p}^\top) \right] \\
&+ rc ~ \tr\left[  \left( \frac{\pd f(\bm{p})}{\pd \bm{p}} \right)  \left( \frac{\pd f(\bm{p})}{\pd \bm{p}^\top} \right) (\diag(\bm{p}) - \bm{p} \bm{p}^\top) \right], \\
&+ \tr \left[ \left( \frac{\pd f(\bm{p})}{\pd \bm{p}} \right) (rc\bm{p} - \bm{1}_{rc})^\top \left( \frac{\pd^2 f(\bm{p})}{\pd \bm{p} \pd \bm{p}^\top} \right) (\diag(\bm{p}) - \bm{p} \bm{p}^\top) \right],
\end{align*}
$\bm{1}_{rc}$ is the $rc \times 1$ vector with all elements equal to one, and $\diag(\bm{p})$ is a diagonal matrix with the elements of $\bm{p}$ on the main diagonal.
\end{thm}
\begin{proof}[Proof of Theorem \ref{thm:mse}]
The MSE of $f(\hat{\bm{p}}^{(\alpha)})$ is expressed as
\begin{align}
\MSE[f(\hat{\bm{p}}^{(\alpha)})] &= \E[(f(\hat{\bm{p}}^{(\alpha)}) - f(\bm{p}))^2] \nonumber \\
&= \E[([f(\hat{\bm{p}}^{(\alpha)}) - f(\hat{\bm{p}})] + [f(\hat{\bm{p}}) - f(\bm{p})])^2]. \label{eq:mse1}
\end{align}
Because $f(\cdot)$ is at least four times differentiable at $\bm{p}$, $f(\hat{\bm{p}})$ is expressed as
\begin{equation}
\label{eq:fhp}
\begin{split}
f(\hat{\bm{p}}) =& f(\bm{p}) + \frac{1}{\sqrt{n}} \left( \frac{\pd f(\bm{p})}{\pd \bm{p}^\top} \right) \bm{u} + \frac{1}{2n} \bm{u}^\top \left( \frac{\pd^2 f(\bm{p})}{\pd \bm{p} \pd \bm{p}^\top} \right) \bm{u} \\
&+ \frac{1}{6n^{3/2}} \left( \sum_{i,j} \sum_{k,l} \sum_{s,t} u_{ij} u_{kl} u_{st} \frac{\pd^3}{\pd p_{ij} \pd p_{kl} \pd p_{st}} \right) f(\bm{p}) + O_p(n^{-2}), 
\end{split}
\end{equation}
where $\bm{u} = (u_{11}, u_{12}, \ldots, u_{1c}, u_{21}, \ldots, u_{rc})^\top$ and $u_{ij} = \sqrt{n}(\hat{p}_{ij} - p_{ij})$.
It should be noted that $\bm{u} = \sqrt{n}(\hat{\bm{p}} - \bm{p}) \overset{d}{\to} N(\bm{0}, \diag(\bm{p}) - \bm{p}\bm{p}^\top)$ as $n\to\infty$.

Additionally, $f(\hat{\bm{p}}^{(\alpha)})$ is expressed as
\begin{align}
f(\hat{\bm{p}}^{(\alpha)}) =& f(\hat{\bm{p}}) + \left(\frac{\pd f(\hat{\bm{p}})}{\pd \hat{\bm{p}}^\top}\right) (\hat{\bm{p}}^{(\alpha)} - \hat{\bm{p}}) + O_p(n^{-2}) \nonumber \\
\begin{split}
=& f(\hat{\bm{p}}) - \frac{\alpha}{n} \left(\frac{\pd f(\bm{p})}{\pd \bm{p}^\top}\right) (rc\bm{p} - \bm{1}_{rc}). \\
&- \frac{\alpha}{n^{3/2}} \bm{u}^\top \left( \frac{\pd^2 f(\bm{p})}{\pd \bm{p} \pd \bm{p}^\top} \right) (rc\bm{p} - \bm{1}_{rc}) - \frac{rc\alpha}{n^{3/2}} \left(\frac{\pd f(\bm{p})}{\pd \bm{p}^\top}\right) \bm{u} + O_p(n^{-2})
\end{split} \label{eq:fhpa}
\end{align}
since
\begin{align*}
\frac{\pd f(\hat{\bm{p}})}{\pd \hat{\bm{p}}^\top} = \frac{\pd f(\bm{p})}{\pd \bm{p}^\top} + \frac{1}{\sqrt{n}} \bm{u}^\top \left( \frac{\pd^2 f(\bm{p})}{\pd \bm{p} \pd \bm{p}^\top} \right) + O_p(n^{-1})
\end{align*}
and
\begin{align*}
\hat{p}_{ij}^{(\alpha)} 
&= \frac{n_{ij}}{n} \left(\frac{1}{1+n^{-1}rc\alpha}\right) + \frac{\alpha}{n} \left(\frac{1}{1+n^{-1}rc\alpha}\right) \\
&= \hat{p}_{ij} \left( 1 - \frac{rc\alpha}{n} + O(n^{-2}) \right) + \frac{\alpha}{n} \left( 1 - \frac{rc\alpha}{n} + O(n^{-2}) \right) \\
&= \hat{p}_{ij} - \frac{rc\alpha}{n} \hat{p}_{ij} + \frac{\alpha}{n} + O_p(n^{-2}) \\
&= \hat{p}_{ij} - \frac{\alpha}{n} (rcp_{ij} - 1) - \frac{rc\alpha}{n^{3/2}} \sqrt{n} (\hat{p}_{ij} - p_{ij}) + O_p(n^{-2}).
\end{align*}

From equations (\ref{eq:mse1}), (\ref{eq:fhp}), and (\ref{eq:fhpa}), we have
\begin{align*}
f(\hat{\bm{p}}^{(\alpha)}) - f(\bm{p}) = \frac{1}{\sqrt{n}}F_1 + \frac{1}{n}(F_{21} - \alpha F_{22}) + \frac{1}{n^{3/2}}(F_{31} - \alpha F_{32} - \alpha F_{33}) + O_p(n^{-2}), 
\end{align*}
where
\begin{gather*}
F_1 = \left( \frac{\pd f(\bm{p})}{\pd \bm{p}^\top} \right) \bm{u}, ~~ F_{21} = \frac{1}{2} \bm{u}^\top \left( \frac{\pd^2 f(\bm{p})}{\pd \bm{p} \pd \bm{p}^\top} \right) \bm{u}, ~~ F_{22} = \left( \frac{\pd f(\bm{p})}{\pd \bm{p}^\top} \right) (rc\bm{p} - \bm{1}_{rc}), \\
F_{31} = \frac{1}{6}\left( \sum_{i,j} \sum_{k,l} \sum_{s,t} u_{ij} u_{kl} u_{st} \frac{\pd^3}{\pd p_{ij} \pd p_{kl} \pd p_{st}} \right) f(\bm{p}), ~~ F_{32} = rc\left( \frac{\pd f(\bm{p})}{\pd \bm{p}^\top} \right) \bm{u}, \\
F_{33} = \bm{u}^\top \left( \frac{\pd^2 f(\bm{p})}{\pd \bm{p} \pd \bm{p}^\top} \right) (rc\bm{p} - \bm{1}_{rc})
\end{gather*}
and then the MSE of $f(\hat{\bm{p}}^{(\alpha)})$ is expressed as
\begin{align*}
\MSE[f(\hat{\bm{p}}^{(\alpha)})] 
=& \frac{1}{n} \E[F_1^2] + \frac{2}{n^{3/2}} \E[F_1(F_{21}-\alpha F_{22})] \\
&+ \frac{1}{n^2} \E[(F_{21}-\alpha F_{22})^2] + \frac{2}{n^2} \E[F_1 (F_{31} - \alpha F_{32} - \alpha F_{33})] + o(n^{-2}) \\
=& \frac{1}{n^2} \left( A_1 \alpha^2  - 2 A_2 \alpha \right) + A_3 + o(n^{-2}), 
\end{align*}
where
\begin{align*}
A_1 =& \tr\left[ \left( \frac{\pd f(\bm{p})}{\pd \bm{p}} \right) \left( \frac{\pd f(\bm{p})}{\pd \bm{p}^\top} \right) (rc\bm{p} - \bm{1}_{rc}) (rc\bm{p} - \bm{1}_{rc})^\top \right], \\
A_2
=& \frac{1}{2} (rc\bm{p} - \bm{1}_{rc})^\top \left( \frac{\pd f(\bm{p})}{\pd \bm{p}} \right) \tr\left[ \left( \frac{\pd^2 f(\bm{p})}{\pd \bm{p} \pd \bm{p}^\top} \right) (\diag(\bm{p}) - \bm{p} \bm{p}^\top) \right] \\
&+ rc ~ \tr\left[  \left( \frac{\pd f(\bm{p})}{\pd \bm{p}} \right)  \left( \frac{\pd f(\bm{p})}{\pd \bm{p}^\top} \right) (\diag(\bm{p}) - \bm{p} \bm{p}^\top) \right], \\
&+ \tr \left[ \left( \frac{\pd f(\bm{p})}{\pd \bm{p}} \right) (rc\bm{p} - \bm{1}_{rc})^\top \left( \frac{\pd^2 f(\bm{p})}{\pd \bm{p} \pd \bm{p}^\top} \right) (\diag(\bm{p}) - \bm{p} \bm{p}^\top) \right], \\
A_3 =& \frac{1}{n} \E[F_1^2] + \frac{2}{n^{3/2}} \E[F_1F_{21}] + \frac{1}{n^2} \E[F_{21}^2] + \frac{2}{n^2} \E[F_1F_{31}],
\end{align*}
since
\begin{align*}
\E[F_1F_{22}] = (rc\bm{p} - \bm{1}_{rc})^\top \left( \frac{\pd f(\bm{p})}{\pd \bm{p}} \right) \left( \frac{\pd f(\bm{p})}{\pd \bm{p}^\top} \right) \E\left[  \bm{u} \right] = 0.
\end{align*}
Note that the terms in $A_3$ are independent of $\alpha$.
\end{proof}

\begin{thm}
\label{thm:param}
The Dirichlet parameter $\alpha$ that asymptotically minimizes the MSE of $f(\hat{\bm{p}}^{(\alpha)})$ is obtained as follows:
\begin{equation*}
\alpha^* = \argmin_{\alpha} \lim_{n\to\infty} n^2 \MSE[f(\hat{\bm{p}}^{(\alpha)})] = \frac{A_2}{A_1}.
\end{equation*}
\end{thm}
\noindent
From Theorem \ref{thm:mse}, it is clear that Theorem \ref{thm:param} holds.

Therefore, using 
\begin{equation}
\label{eq:minA}
\hat{\alpha}^* = \frac{\hat{A}_2}{\hat{A}_1},
\end{equation}
where $\hat{A}_1$ and $\hat{A}_2$ denote $A_1$ and $A_2$ with $\bm{p}$ replaced by $\hat{\bm{p}}$, respectively. 
we propose $f(\hat{\bm{p}}^{(\hat{\alpha}^*)})$ as an estimator of the index $f(\bm{p})$.

%=sec:experiments=========================================================
\section{Numerical Experiments}
\label{sec:experiments}
This section shows that the proposed estimator $f(\hat{\bm{p}}^{(\hat{\alpha}^*)})$ can reduce the bias and MSE more than the estimator with sample proportions $f(\hat{\bm{p}})$ and the improved estimator (e.g., \citealp{tomizawa2007improved}) in the numerical experiments.
We also show that the Dirichlet parameter $\hat{\alpha}^*$ chosen by the proposed method (\ref{eq:minA}) is more suitable for improving the accuracy for estimating indexes than other methods of choosing the Dirichlet parameter (e.g., $\alpha=1$ (uniform prior), $\alpha=1/2$ (Jeffreys prior), and \cite{fienberg1973simultaneous}'s method (FHM)).
The bias and MSE based on the numerical experiments are calculated as 
\begin{align*}
{\rm Bias} = \frac{1}{S} \sum_{i = 1}^S f_{i}(\bm{p}^*) - f(\bm{p}), ~~ {\rm MSE} = \frac{1}{S} \sum_{i=1}^S (f_{i}(\bm{p}^*) - f(\bm{p}))^2,
\end{align*}
where $S$ is the number of times a multinomial random number is generated, and $f_{i}(\bm{p}^*)$ is the estimated value of $f(\bm{p})$ at the $i$th multinomial random number.
These numerical experiments are performed using the programming language \textbf{R} (\citealp{r2022r}).

\subsection{Numerical Experiment for Index in Two-Way Contingency Tables}
\label{sec:gcc}
First, we consider the generalized Cram\'{e}r's coefficient in $r \times c$ contingency tables.
\cite{tomizawa2004generalization} proposed the generalized Cram\'{e}r's coefficient where the column variable is the explanatory variable and the row variable is the response variable as follows:
\begin{align*}
f(\bm{p}) = V^{(\lambda)} = \frac{I^{(\lambda)}(\{p_{ij}\};\{p_{i\cdot}p_{\cdot j}\})}{K^{(\lambda)}} ~~ \mbox{for $\lambda \geq 0$},
\end{align*}
where
\begin{gather*}
I^{(\lambda)}(\{p_{ij}\};\{p_{i\cdot}p_{\cdot j}\}) = \frac{1}{\lambda(\lambda+1)} \sum_{i,j} p_{ij} \left[ \left(\frac{p_{ij}}{p_{i \cdot}p_{\cdot j}}\right)^\lambda - 1 \right], \\
p_{i \cdot} = \sum_{j} p_{ij}, ~~ p_{\cdot j} = \sum_{i} p_{ij}, ~~ K^{(\lambda)} = \frac{1}{\lambda(\lambda+1)} \left( \sum_{i} p_{i \cdot}^{1-\lambda} - 1 \right),
\end{gather*}
and the value at $\lambda = 0$ is taken as the continuous limit as $\lambda \to 0$.
Note that $I^{(\lambda)}(\cdot;\cdot)$ is the power-divergence between two distributions $\{p_{ij}\}$ and $\{p_{i\cdot}p_{\cdot j}\}$, including the Kullback Leibler information ($\lambda=0$) and one-half of the Pearson chi-squared type discrepancy ($\lambda=1$), and the real number $\lambda$ is chosen by the user.
In this numerical experiment, we consider the case of $\lambda = 1$ for simplicity.

Suppose that $4 \times 5$ contingency tables are generated 10000 times by a multinomial random number based on the structures of probabilities in Tables \ref{tb:ne1}a, \ref{tb:ne1}b, and \ref{tb:ne1}c.
These probability tables are constructed so that the values are at both ends of the range of the generalized Cram\'{e}r's coefficient and in the middle of the range, in fact, these values for Tables \ref{tb:ne1}a, \ref{tb:ne1}b, and \ref{tb:ne1}c are $0.091$, $0.486$, and $0.819$, respectively.

\begin{table}[h]
\caption{The $4 \times 5$ structures of probabilities to generate a multinomial random number}
\begin{center}
\scalebox{0.63}{
\begin{tabular}{ccccc}
\hline
\begin{tabular}{rrrrr}
\multicolumn{5}{l}{(a)} \\ \hline
$0.048$&$0.055$&$0.105$&$0.023$&$0.018$\tabularnewline
$0.032$&$0.061$&$0.035$&$0.018$&$0.098$\tabularnewline
$0.055$&$0.131$&$0.016$&$0.082$&$0.054$\tabularnewline
$0.029$&$0.012$&$0.032$&$0.033$&$0.063$\tabularnewline
\hline
\end{tabular}
&&
\begin{tabular}{rrrrr}
\multicolumn{5}{l}{(b)} \\ \hline
$0.154$&$0.013$&$0.021$&$0.018$&$0.145$\tabularnewline
$0.017$&$0.017$&$0.159$&$0.015$&$0.012$\tabularnewline
$0.015$&$0.157$&$0.011$&$0.017$&$0.018$\tabularnewline
$0.013$&$0.011$&$0.013$&$0.163$&$0.011$\tabularnewline
\hline
\end{tabular}
&&
\begin{tabular}{rrrrr}
\multicolumn{5}{l}{(c)} \\ \hline
$0.185$&$0.006$&$0.003$&$0.006$&$0.182$\tabularnewline
$0.004$&$0.003$&$0.188$&$0.005$&$0.004$\tabularnewline
$0.005$&$0.187$&$0.004$&$0.007$&$0.003$\tabularnewline
$0.006$&$0.004$&$0.005$&$0.190$&$0.003$\tabularnewline
\hline
\end{tabular}
\end{tabular}
}
\end{center}
\label{tb:ne1}
\end{table}%

\begin{figure}[H]
\begin{center}
\includegraphics[width=\columnwidth]{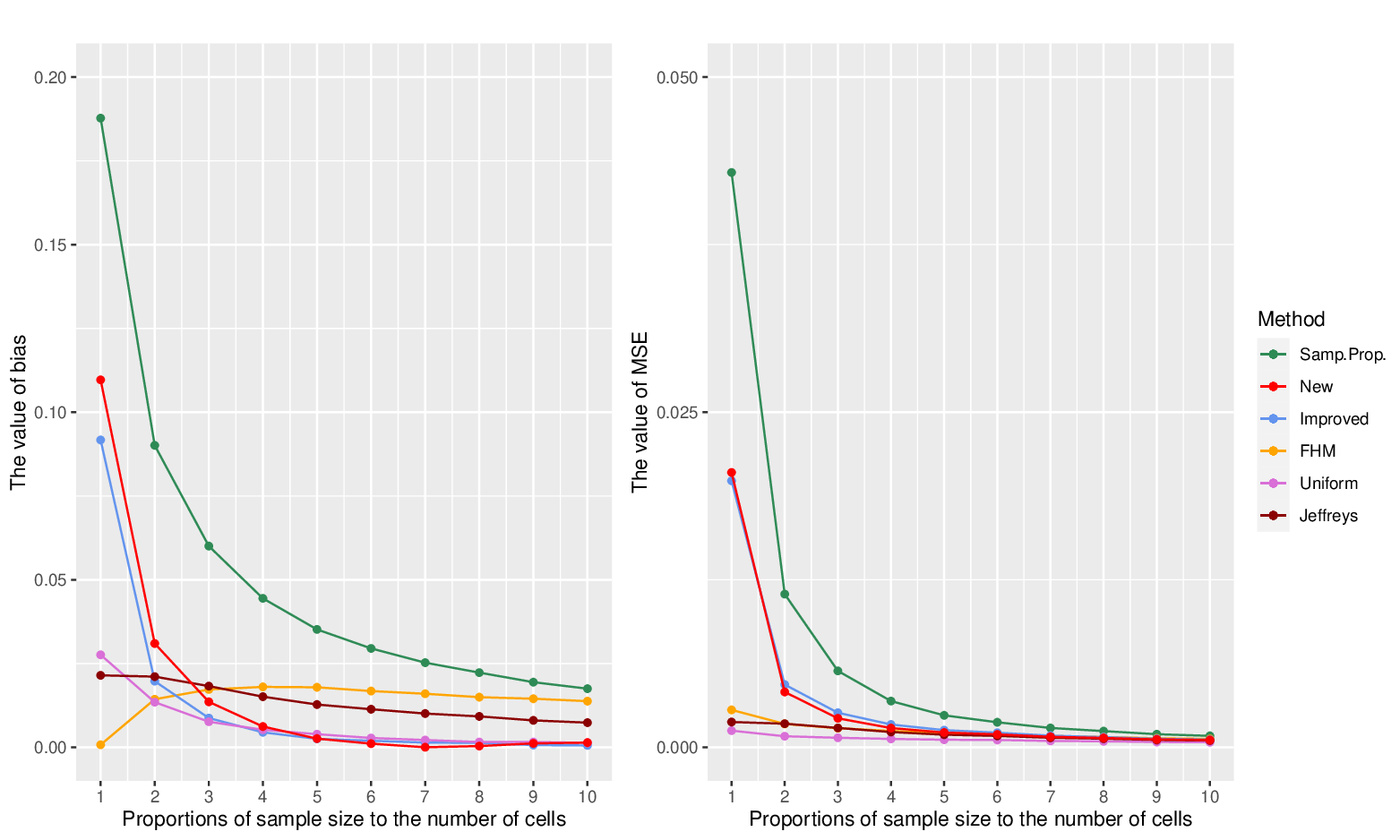}
\caption{The absolute values of the bias and the values of MSE for several estimators of the generalized Cram\'{e}r's coefficient with Table \ref{tb:ne1}a when $\gamma = 1, 2, \ldots, 10$, where $\gamma$ is the proportion of sample size to the number of cells 
(Samp.Prop. (green line): plug-in estimator with the sample proportions; New (red line): the proposed estimator; Improved (light blue line): Improved estimator; FHM (yellow line): plug-in estimator with the posterior means of the cell probabilities using the Dirichlet parameter chosen by \cite{fienberg1973simultaneous}'s method; Uniform (pink line): plug-in estimator with the posterior means of the cell probabilities using the uniform prior; Jeffreys (brown line): plug-in estimator with the posterior means of the cell probabilities using Jeffreys prior)}
\label{fig:ne1a}
\end{center}
\end{figure}

\begin{figure}[H]
\begin{center}
\includegraphics[width=\columnwidth]{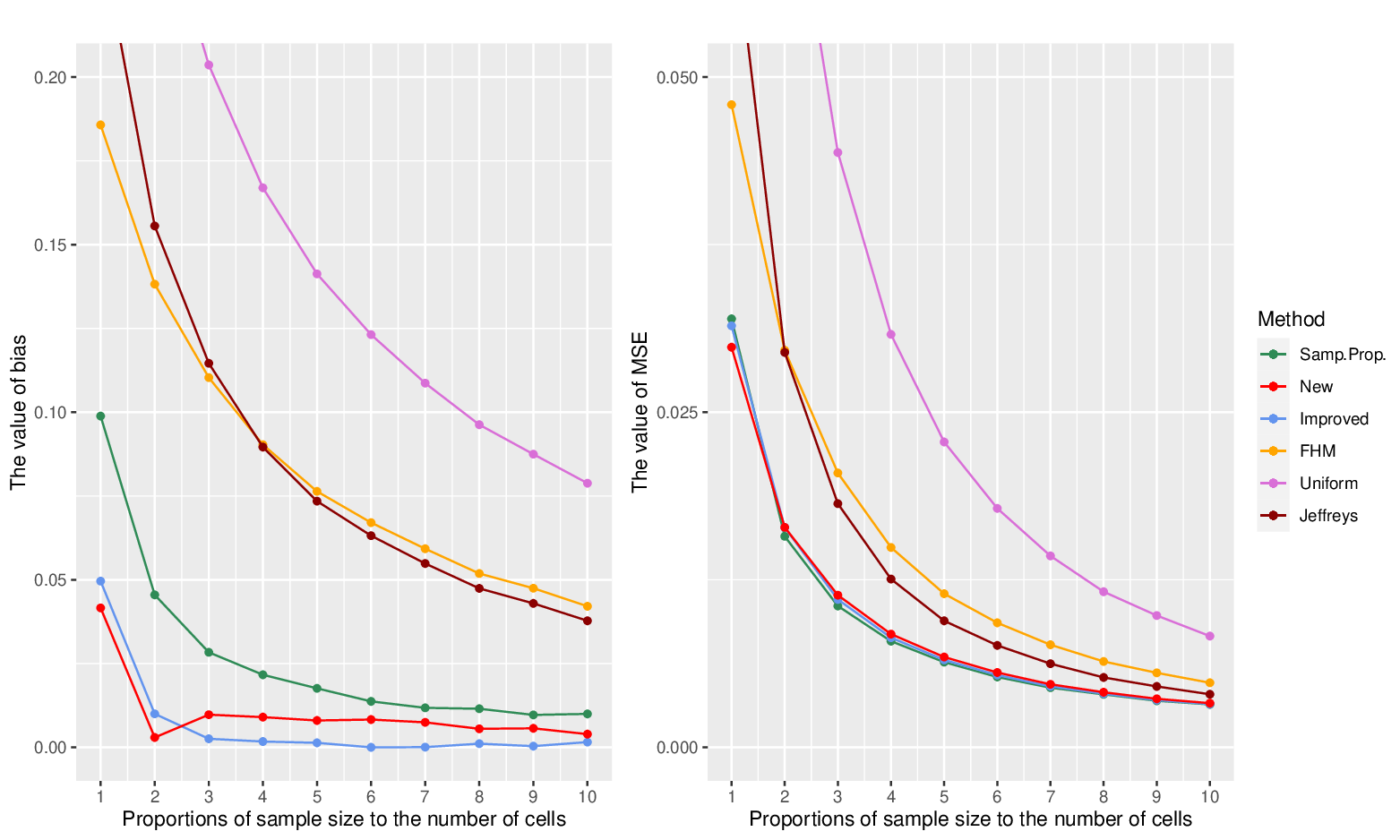}
\caption{The absolute values of the bias and the values of MSE for several estimators of the generalized Cram\'{e}r's coefficient with Table \ref{tb:ne1}b when $\gamma = 1, 2, \ldots, 10$, where $\gamma$ is the proportion of sample size to the number of cells
(Samp.Prop. (green line): plug-in estimator with the sample proportions; New (red line): the proposed estimator; Improved (light blue line): Improved estimator; FHM (yellow line): plug-in estimator with the posterior means of the cell probabilities using the Dirichlet parameter chosen by \cite{fienberg1973simultaneous}'s method; Uniform (pink line): plug-in estimator with the posterior means of the cell probabilities using the uniform prior; Jeffreys (brown line): plug-in estimator with the posterior means of the cell probabilities using Jeffreys prior)}
\label{fig:ne1b}
\end{center}
\end{figure}

\begin{figure}[H]
\begin{center}
\includegraphics[width=\columnwidth]{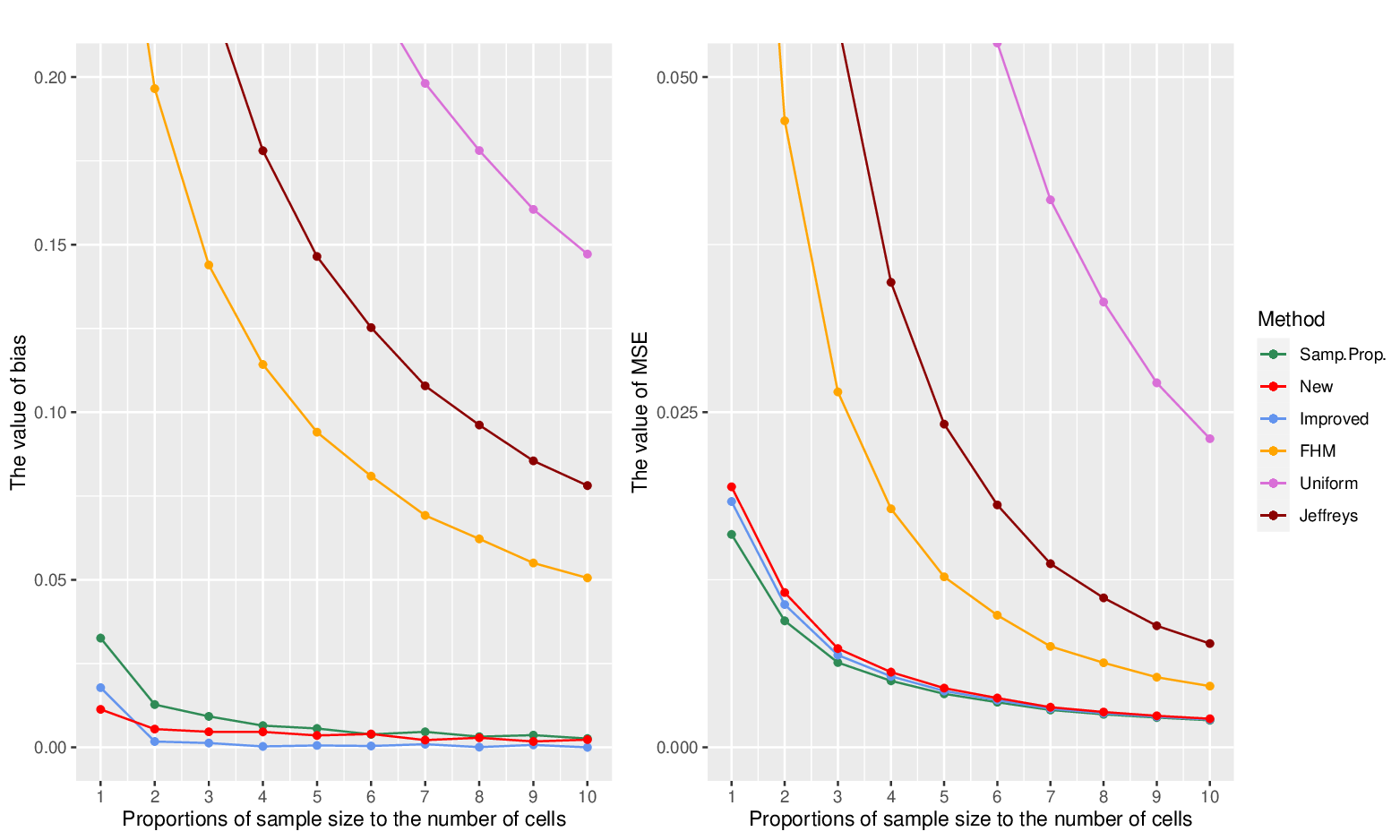}
\caption{The absolute values of the bias and the values of MSE for several estimators of the generalized Cram\'{e}r's coefficient with Table \ref{tb:ne1}c when $\gamma = 1, 2, \ldots, 10$, where $\gamma$ is the proportion of sample size to the number of cells
(Samp.Prop. (green line): plug-in estimator with the sample proportions; New (red line): the proposed estimator; Improved (light blue line): Improved estimator; FHM (yellow line): plug-in estimator with the posterior means of the cell probabilities using the Dirichlet parameter chosen by \cite{fienberg1973simultaneous}'s method; Uniform (pink line): plug-in estimator with the posterior means of the cell probabilities using the uniform prior; Jeffreys (brown line): plug-in estimator with the posterior means of the cell probabilities using Jeffreys prior)}
\label{fig:ne1c}
\end{center}
\end{figure}

Figures \ref{fig:ne1a}, \ref{fig:ne1b}, and \ref{fig:ne1c} represent the absolute value of bias and MSE for several estimators of the generalized Cram\'{e}r's coefficient with Table \ref{tb:ne1} when $\gamma = 1, 2, \ldots, 10$, where $\gamma$ is the proportion of sample size to the number of cells.

As can be seen from Figures \ref{fig:ne1a} and \ref{fig:ne1b}, the estimator with the sample proportions $f(\hat{\bm{p}})$ (green line) has the large values of bias and MSE overall, while the values of bias and MSE are smaller in Figure \ref{fig:ne1c}, because Table \ref{tb:ne1}c has many cells with probabilities close to zero and the estimation accuracy of the sample proportions is good.
Our proposed estimator $f(\hat{\bm{p}}^{(\hat{\alpha}^*)})$ (red line) has significantly improved the estimation accuracy compared to the conventional estimator $f(\hat{\bm{p}})$ (green line) in the situations of Tables \ref{tb:ne1}a and \ref{tb:ne1}b, and has the same estimation accuracy as $f(\hat{\bm{p}})$ in the situation of Table \ref{tb:ne1}c, where $f(\hat{\bm{p}})$ has the good estimation accuracy.
It is also found that $f(\hat{\bm{p}}^{(\hat{\alpha}^*)})$ has the smaller values of the MSE than the improved estimator with bias correction (light blue line), although slightly.
On the other hand, the improved estimator has the smaller values of the bias, but $f(\hat{\bm{p}}^{(\hat{\alpha}^*)})$ has the smaller bias value when $\gamma$ is small.

Comparing the estimators of the index using the Bayesian estimators of the cell probabilities with each other, in Figure \ref{fig:ne1a}, the estimators with the posterior means of the cell probabilities using the uniform prior $f(\hat{\bm{p}}^{(1)})$ (pink line) and Jeffreys prior $f(\hat{\bm{p}}^{(1/2)})$ (brown line), and the estimator with the posterior means of the cell probabilities using the Dirichlet parameter chosen by \cite{fienberg1973simultaneous}'s method (yellow line) show almost the same estimation accuracy, and the values of bias and MSE are small, which may be due to the appropriate prior information.
As the value of $\gamma$ increases, i.e., as the number of samples increases, the estimation accuracy of our proposed estimator $f(\hat{\bm{p}}^{(\hat{\alpha}^*)})$ becomes equal to or better than those estimators.
Whereas, in Figures \ref{fig:ne1b}, and \ref{fig:ne1c}, the values of the bias and MSE of $f(\hat{\bm{p}}^{(1)})$ (pink line), $f(\hat{\bm{p}}^{(1/2)})$ (brown line), and the estimator with the Dirichlet parameter chosen by \cite{fienberg1973simultaneous}'s method (yellow line) are much larger than those of $f(\hat{\bm{p}})$.
However, as mentioned above, our proposed estimator $f(\hat{\bm{p}}^{(\hat{\alpha}^*)})$ shows better estimation accuracy than $f(\hat{\bm{p}})$ in Figure \ref{fig:ne1b} and the same accuracy as $f(\hat{\bm{p}})$ in Figure \ref{fig:ne1c}, indicating that it is the stable estimation method in all situations in Table \ref{tb:ne1}.

\subsection{Numerical Experiment for Index in Square Contingency Tables}
\label{sec:s}
Next, we consider a index to represent the degree of departure from the symmetry model in square contingency tables.
\cite{tomizawa1998power} proposed the index to represent the degree of departure from the symmetry model as follows.
\begin{align*}
f(\bm{p}) = \Phi^{(\lambda)} = \sum_{i<j} (p_{ij}^* + p_{ji}^*) \phi_{ij}^{(\lambda)} ~~ \mbox{for $\lambda > -1$},
\end{align*}
where
\begin{gather*}
\phi_{ij}^{(\lambda)} = 1 - \frac{\lambda 2^\lambda}{2^\lambda - 1} H_{ij}^{(\lambda)}, ~~ H_{ij}^{(\lambda)} = \frac{1}{\lambda} \left[ 1 - (p_{ij}^c)^{\lambda+1} - (p_{ji}^c)^{\lambda+1} \right], \\
p_{ij}^c = \frac{p_{ij}}{p_{ij} + p_{ji}}, ~~ p_{ij}^* = \frac{p_{ij}}{\delta}, ~~ \delta = \sum_{i \neq j} p_{ij},
\end{gather*}
and the value at $\lambda = 0$ is taken as the continuous limit as $\lambda \to 0$.
Note that $H_{ij}^{(\lambda)}$ is the \cite{patil1982diversity}'s diversity index of degree $\lambda$, including the Shannon entropy ($\lambda = 0$), and the real number $\lambda$ is chosen by the user.
In this numerical experiment, we consider the case of $\lambda = 1$ for simplicity.

Suppose that $4 \times 4$ square contingency tables are generated 10000 times by a multinomial random number based on the structures of probabilities in Tables \ref{tb:ne2}a, \ref{tb:ne2}b, and \ref{tb:ne2}c.
These probability tables are constructed so that the values are at both ends of the range of the index for symmetry and in the middle of the range, in fact, these values for Tables \ref{tb:ne2}a, \ref{tb:ne2}b, and \ref{tb:ne2}c are $0.099$, $0.473$, and $0.800$, respectively.
Additionally, we assume that contingency tables whose rows and columns consist of the same classification may have larger probabilities of the main diagonal cells.

\begin{table}[h]
\caption{The $4 \times 4$ structures of probabilities to generate a multinomial random number}
\begin{center}
\scalebox{0.75}{
\begin{tabular}{ccccc}
\hline
\begin{tabular}{rrrr}
\multicolumn{4}{l}{(a)} \\ \hline
$0.100$&$0.060$&$0.038$&$0.071$\tabularnewline
$0.038$&$0.100$&$0.061$&$0.026$\tabularnewline
$0.068$&$0.051$&$0.100$&$0.031$\tabularnewline
$0.029$&$0.066$&$0.061$&$0.100$\tabularnewline
\hline
\end{tabular}
&&
\begin{tabular}{rrrr}
\multicolumn{4}{l}{(b)} \\ \hline
$0.100$&$0.018$&$0.012$&$0.007$\tabularnewline
$0.094$&$0.100$&$0.021$&$0.014$\tabularnewline
$0.082$&$0.089$&$0.100$&$0.023$\tabularnewline
$0.071$&$0.081$&$0.088$&$0.100$\tabularnewline
\hline
\end{tabular}
&&
\begin{tabular}{rrrr}
\multicolumn{4}{l}{(c)} \\ \hline
$0.100$&$0.089$&$0.004$&$0.102$\tabularnewline
$0.005$&$0.100$&$0.094$&$0.007$\tabularnewline
$0.084$&$0.002$&$0.100$&$0.111$\tabularnewline
$0.009$&$0.088$&$0.005$&$0.100$\tabularnewline
\hline
\end{tabular}
\end{tabular}
}
\end{center}
\label{tb:ne2}
\end{table}%

\begin{figure}[H]
\begin{center}
\includegraphics[width=\columnwidth]{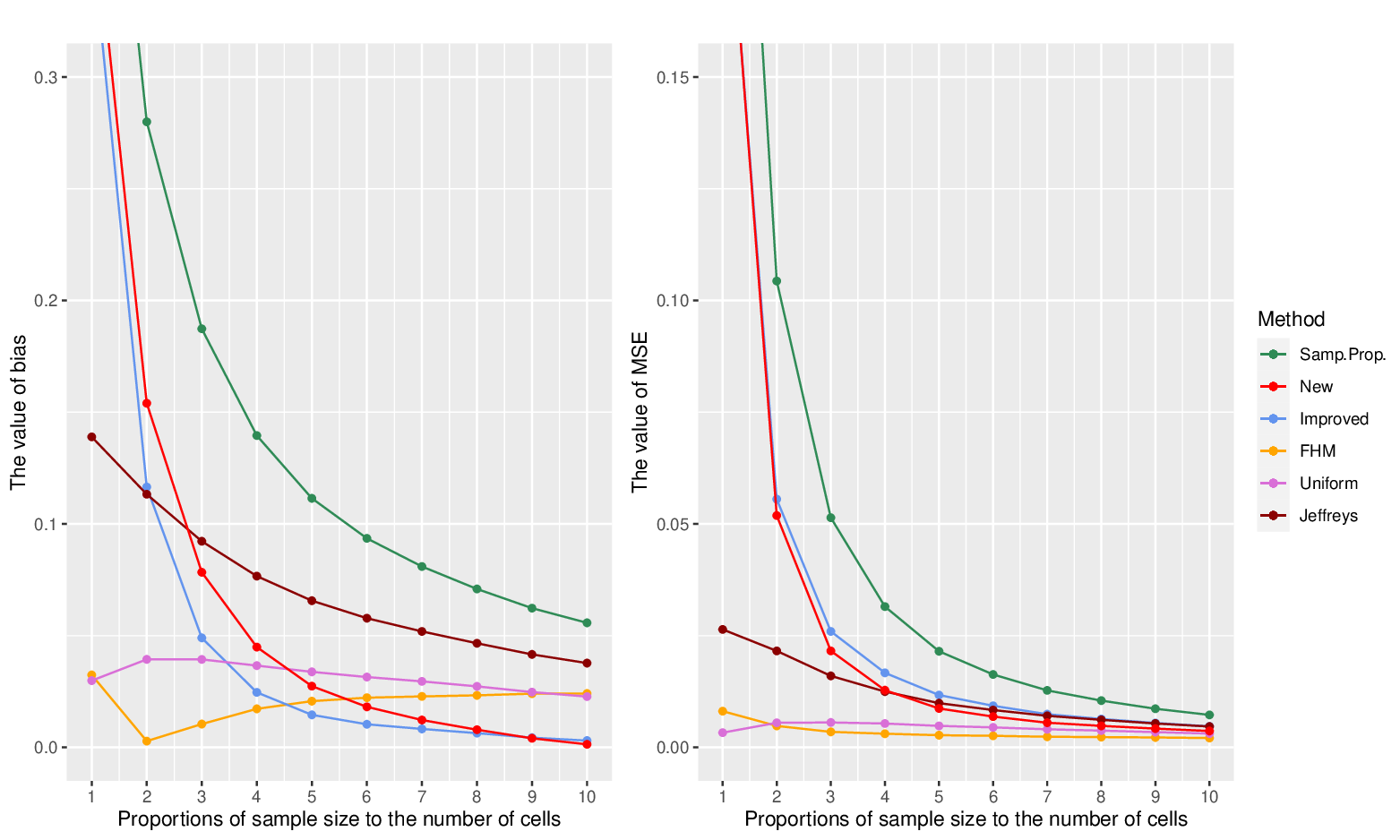}
\caption{The absolute values of the bias and the values of MSE for several estimators of the index to represent the degree of departure from symmetry with Table \ref{tb:ne2}a when $\gamma = 1, 2, \ldots, 10$, where $\gamma$ is the proportion of sample size to the number of cells
(Samp.Prop. (green line): plug-in estimator with the sample proportions; New (red line): the proposed estimator; Improved (light blue line): Improved estimator; FHM (yellow line): plug-in estimator with the posterior means of the cell probabilities using the Dirichlet parameter chosen by \cite{fienberg1973simultaneous}'s method; Uniform (pink line): plug-in estimator with the posterior means of the cell probabilities using the uniform prior; Jeffreys (brown line): plug-in estimator with the posterior means of the cell probabilities using Jeffreys prior)}
\label{fig:ne2a}
\end{center}
\end{figure}

\begin{figure}[H]
\begin{center}
\includegraphics[width=\columnwidth]{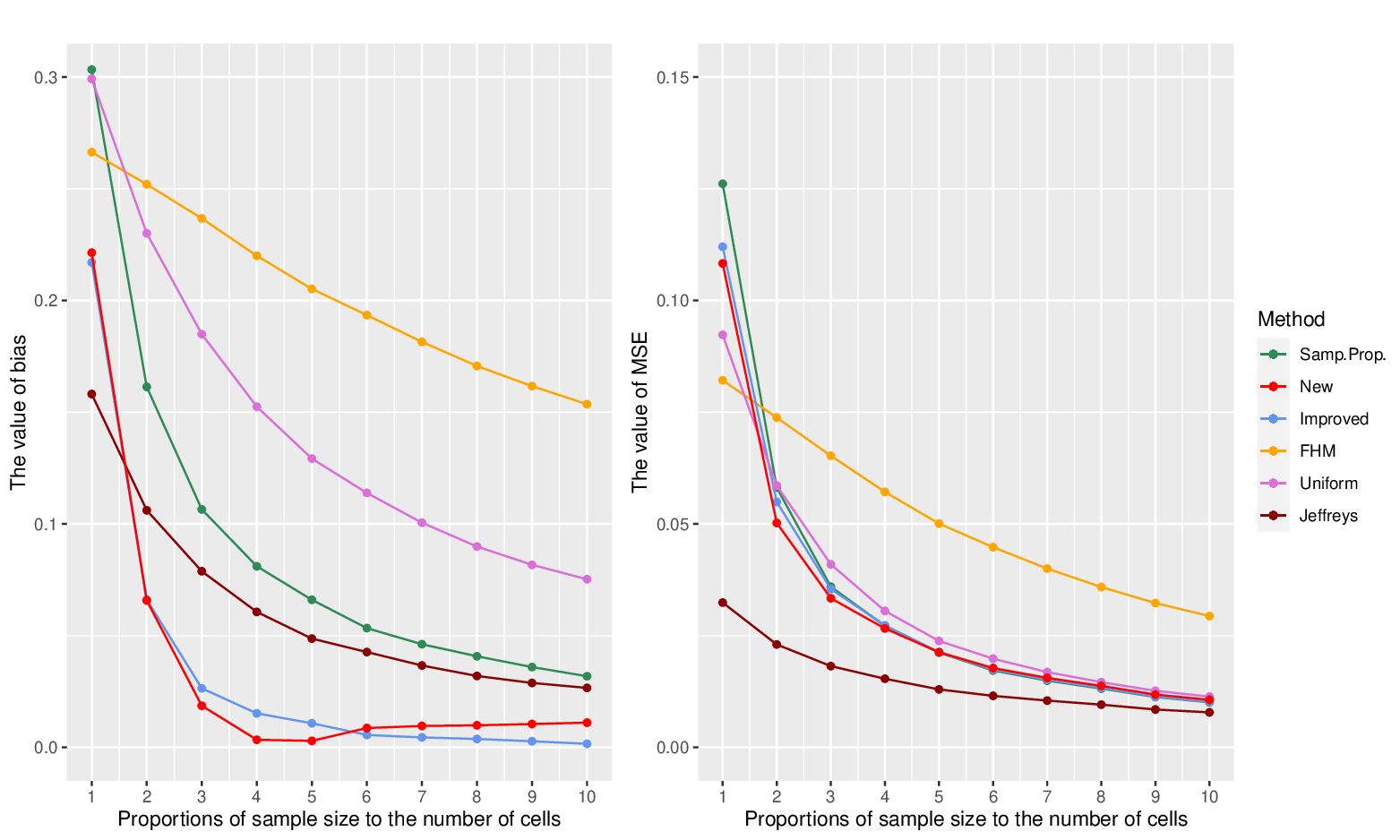}
\caption{The absolute values of the bias and the values of MSE for several estimators of the index to represent the degree of departure from symmetry with Table \ref{tb:ne2}b when $\gamma = 1, 2, \ldots, 10$, where $\gamma$ is the proportion of sample size to the number of cells
(Samp.Prop. (green line): plug-in estimator with the sample proportions; New (red line): the proposed estimator; Improved (light blue line): Improved estimator; FHM (yellow line): plug-in estimator with the posterior means of the cell probabilities using the Dirichlet parameter chosen by \cite{fienberg1973simultaneous}'s method; Uniform (pink line): plug-in estimator with the posterior means of the cell probabilities using the uniform prior; Jeffreys (brown line): plug-in estimator with the posterior means of the cell probabilities using Jeffreys prior)}
\label{fig:ne2b}
\end{center}
\end{figure}

\begin{figure}[H]
\begin{center}
\includegraphics[width=\columnwidth]{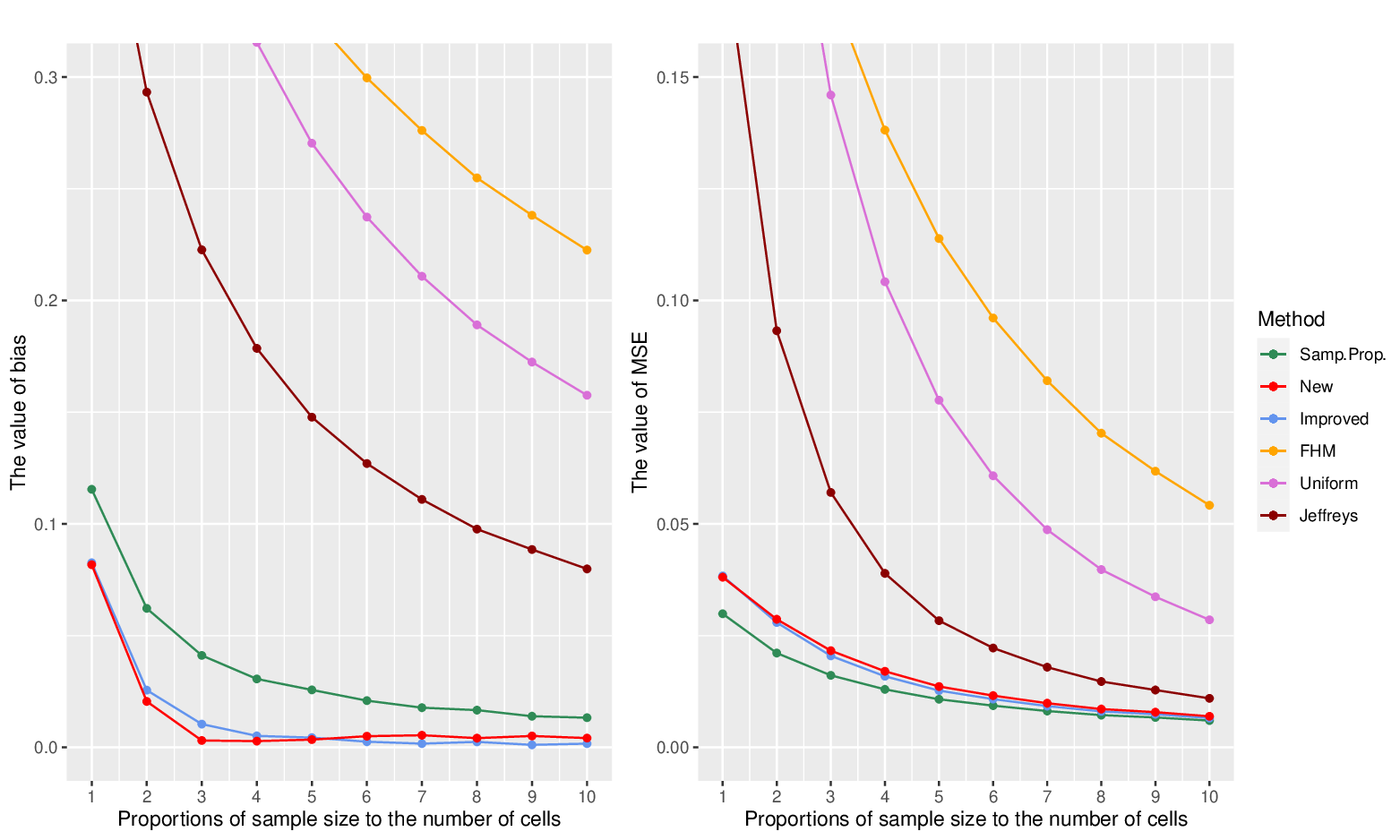}
\caption{The absolute values of the bias and the values of MSE for several estimators of the index to represent the degree of departure from symmetry with Table \ref{tb:ne2}c when $\gamma = 1, 2, \ldots, 10$, where $\gamma$ is the proportion of sample size to the number of cells
(Samp.Prop. (green line): plug-in estimator with the sample proportions; New (red line): the proposed estimator; Improved (light blue line): Improved estimator; FHM (yellow line): plug-in estimator with the posterior means of the cell probabilities using the Dirichlet parameter chosen by \cite{fienberg1973simultaneous}'s method; Uniform (pink line): plug-in estimator with the posterior means of the cell probabilities using the uniform prior; Jeffreys (brown line): plug-in estimator with the posterior means of the cell probabilities using Jeffreys prior)}
\label{fig:ne2c}
\end{center}
\end{figure}

Figures \ref{fig:ne2a}, \ref{fig:ne2b}, and \ref{fig:ne2c} represent the absolute value of bias and MSE for several estimators of the index to represent the degree of departure from symmetry with Table \ref{tb:ne2} when $\gamma = 1, 2, \ldots, 10$, where $\gamma$ is the proportion of sample size to the number of cells.

Similar to the numerical experiments in Section \ref{sec:gcc}, the estimator with the sample proportions $f(\hat{\bm{p}})$ (green line) has poor estimation accuracy in the case of Tables \ref{tb:ne2}a and \ref{tb:ne2}b, which have few cells with zero probability (Figures \ref{fig:ne2a} and \ref{fig:ne2b}), but good accuracy in the case of Table \ref{tb:ne2}c, which has many cells with zero probability (Figure \ref{fig:ne2c}).
In this numerical experimental setting, our proposed estimator $f(\hat{\bm{p}}^{(\hat{\alpha}^*)})$ (red line) also has significantly improved the estimation accuracy compared to the conventional estimator $f(\hat{\bm{p}})$ (green line) in the situations of Tables \ref{tb:ne2}a and \ref{tb:ne2}b, and has the same estimation accuracy as $f(\hat{\bm{p}})$ in the situation of Table \ref{tb:ne2}c, where $f(\hat{\bm{p}})$ has the good estimation accuracy.
It is also found that $f(\hat{\bm{p}}^{(\hat{\alpha}^*)})$ has the smaller values of the MSE than the improved estimator with bias correction (light blue line), although slightly.
On the other hand, the improved estimator has the smaller values of the bias, but $f(\hat{\bm{p}}^{(\hat{\alpha}^*)})$ has the smaller bias value when $\gamma$ is small.

Comparing the estimators of the index using the Bayesian estimators of the cell probabilities with each other, in Figure \ref{fig:ne2a}, the estimators with the posterior means of the cell probabilities using the uniform prior $f(\hat{\bm{p}}^{(1)})$ (pink line) and Jeffreys prior $f(\hat{\bm{p}}^{(1/2)})$ (brown line), and the estimator with the posterior means of the cell probabilities using the Dirichlet parameter chosen by \cite{fienberg1973simultaneous}'s method (yellow line) show almost the same estimation accuracy, and the values of bias and MSE are small, which may be due to the appropriate prior information.
As the value of $\gamma$ increases, i.e., as the number of samples increases, the estimation accuracy of our proposed estimator $f(\hat{\bm{p}}^{(\hat{\alpha}^*)})$ becomes equal to or better than those estimators.
Whereas, in Figures \ref{fig:ne2b}, and \ref{fig:ne2c}, the values of the bias and MSE of $f(\hat{\bm{p}}^{(1)})$ (pink line) and the estimator with the Dirichlet parameter chosen by \cite{fienberg1973simultaneous}'s method (yellow line) are much larger than those of $f(\hat{\bm{p}})$.
In Figure \ref{fig:ne2b}, the values of MSE of $f(\hat{\bm{p}}^{(1/2)})$ (brown line) are smaller, but the values of bias are larger, and in Figure \ref{fig:ne2c}, the values of the bias and MSE of $f(\hat{\bm{p}}^{(1/2)})$ are much larger.
However, as mentioned above, our proposed estimator $f(\hat{\bm{p}}^{(\hat{\alpha}^*)})$ shows better estimation accuracy than $f(\hat{\bm{p}})$ in Figure \ref{fig:ne2b} and the same accuracy as $f(\hat{\bm{p}})$ in Figure \ref{fig:ne2c}, indicating that it is the stable estimation method in all situations in Table \ref{tb:ne2}.

\begin{rem}
As mentioned in Section \ref{sec:intro}, the range of value for the improved estimator does not equal the range of value for the index.
This problem could be addressed by matching the range of value for the index with that of the improved estimator using a transformation based on the logit function or some other functions.
Here, we describe the accuracy of the improved estimator with the logit transformation through the setting of numerical experiments in Sections \ref{sec:gcc} and \ref{sec:s}.
\\
\indent
Tables \ref{tb:ne1_logit} and \ref{tb:ne2_logit} show the absolute values of the bias and the values of MSE for the improved estimators with the logit transformation of \cite{tomizawa2004generalization}'s and \cite{tomizawa1998power}'s indexes, $V^{(\lambda)}$ and $\Phi^{(\lambda)}$, when $\gamma = 1, 2, \ldots, 10$, where $\gamma$ is the proportion of sample size to the number of cells, respectively.
From these results, it is important to note that the values of the bias and MSE do not approach zero even when the value of $\gamma$ increases, i.e., when the sample size increases given the fixed number of cells.
While the improved estimator is the asymptotically unbiased estimator and thus the value of the bias approaches zero as the sample size increases, the improved estimator with the logit transformation is not the asymptotically unbiased estimator, as shown in these numerical experiments.
The improved estimator, which is still the asymptotically unbiased estimator even after a certain transformation, is a subject for future work.
\end{rem}

\begin{table}[h]
\caption{
The absolute values of the bias and the values of MSE for the improved estimator with the logit transformation of \cite{tomizawa2004generalization}'s index $V^{(\lambda)}$ in Tables \ref{tb:ne1}a, \ref{tb:ne1}b, \ref{tb:ne1}c, \ref{tb:ne1add}d, and \ref{tb:ne1add}e when $\gamma = 1, 2, \ldots, 10$, where $\gamma$ is the proportion of sample size to the number of cells
}
\begin{center}
\scalebox{0.9}{
\begin{tabular}{l rrrrrrrrrr} \hline
 & \multicolumn{10}{c}{$\gamma$} \\ \cline{2-11}
 & 1 & 2 & 3 & 4 & 5 & 6 & 7 & 8 & 9 & 10 \\ \hline
\textbf{Bias} & \multicolumn{10}{c}{} \\ 
Table \ref{tb:ne1}a & 0.454 & 0.436 & 0.434 & 0.433 & 0.432 & 0.432 & 0.432 & 0.432 & 0.432 & 0.432 \\ 
Table \ref{tb:ne1}b & 0.144 & 0.135 & 0.134 & 0.133 & 0.134 & 0.133 & 0.134 & 0.133 & 0.133 & 0.133 \\ 
Table \ref{tb:ne1}c & 0.122 & 0.125 & 0.125 & 0.125 & 0.125 & 0.125 & 0.125 & 0.125 & 0.125 & 0.125 \\
\multicolumn{11}{c}{} \\
\textbf{MSE} & \multicolumn{10}{c}{} \\ 
Table \ref{tb:ne1}a & 0.207 & 0.190 & 0.188 & 0.187 & 0.187 & 0.187 & 0.187 & 0.187 & 0.186 & 0.186 \\ 
Table \ref{tb:ne1}b & 0.022 & 0.019 & 0.019 & 0.018 & 0.018 & 0.018 & 0.018 & 0.018 & 0.018 & 0.018 \\ 
Table \ref{tb:ne1}c & 0.016 & 0.016 & 0.016 & 0.016 & 0.016 & 0.016 & 0.016 & 0.016 & 0.016 & 0.016 \\ \hline 
\end{tabular}
}
\end{center}
\label{tb:ne1_logit}
\end{table}%

\begin{table}[h]
\caption{
The absolute values of the bias and the values of MSE for the improved estimator with the logit transformation of \cite{tomizawa1998power}'s index $\Phi^{(\lambda)}$ in Tables \ref{tb:ne2}a, \ref{tb:ne2}b, \ref{tb:ne2}c, and \ref{tb:ne2add} when $\gamma = 1, 2, \ldots, 10$, where $\gamma$ is the proportion of sample size to the number of cells
}
\begin{center}
\scalebox{0.9}{
\begin{tabular}{l rrrrrrrrrr} \hline
 & \multicolumn{10}{c}{$\gamma$} \\ \cline{2-11}
 & 1 & 2 & 3 & 4 & 5 & 6 & 7 & 8 & 9 & 10 \\ \hline
\textbf{Bias} & \multicolumn{10}{c}{} \\ 
Table \ref{tb:ne2}a & 0.514 & 0.456 & 0.439 & 0.432 & 0.430 & 0.428 & 0.428 & 0.428 & 0.427 & 0.427 \\ 
Table \ref{tb:ne2}b & 0.191 & 0.157 & 0.148 & 0.147 & 0.145 & 0.144 & 0.144 & 0.144 & 0.143 & 0.143 \\ 
Table \ref{tb:ne2}c & 0.094 & 0.106 & 0.108 & 0.109 & 0.109 & 0.110 & 0.110 & 0.110 & 0.110 & 0.110 \\
\multicolumn{11}{c}{} \\
\textbf{MSE} & \multicolumn{10}{c}{} \\ 
Table \ref{tb:ne2}a & 0.268 & 0.211 & 0.194 & 0.188 & 0.186 & 0.184 & 0.184 & 0.183 & 0.183 & 0.183 \\ 
Table \ref{tb:ne2}b & 0.040 & 0.027 & 0.024 & 0.023 & 0.022 & 0.022 & 0.021 & 0.021 & 0.021 & 0.021 \\ 
Table \ref{tb:ne2}c & 0.010 & 0.013 & 0.013 & 0.013 & 0.013 & 0.013 & 0.013 & 0.013 & 0.012 & 0.012 \\ \hline 
\end{tabular}
}
\end{center}
\label{tb:ne2_logit}
\end{table}%

\begin{rem}
Here we show how much the improved estimates does not fall within the interval $[0,1]$ in the settings of the numerical experiments described above and in some additional probability structures.
\\
\\
\noindent
\textbf{Case 1: \cite{tomizawa2004generalization}'s index}
\\
Consider the probability tables in Tables \ref{tb:ne1add}d and \ref{tb:ne1add}e in addition to those in Tables \ref{tb:ne1}a, \ref{tb:ne1}b, and \ref{tb:ne1}c.
The values of the index $V^{(\lambda)}$ in Tables \ref{tb:ne1add}d and \ref{tb:ne1add}e are 0.006 and 0.971, respectively.
The closer the value of the index is to the boundary of its range, the more likely the improved estimator is to be in the outside of the range.
In fact, as shown in Table \ref{tb:ne1r_add}, the number of times that the improved estimate is in the outside of the range of the index for the five probability structures for 10000 simulations is the highest for Table \ref{tb:ne1add}d at about 30\%, followed by 10\% for Table \ref{tb:ne1add}e.
This occurs more frequently for smaller values of $\gamma$, the proportion of the number of samples to the number of cells.
\\
\\
\\
\noindent
\textbf{Case 2: \cite{tomizawa1998power}'s index}
\\
Consider the probability tables in Table \ref{tb:ne2add} in addition to those in Tables \ref{tb:ne2}a, \ref{tb:ne2}b, and \ref{tb:ne2}c.
The value of the index $\Phi^{(\lambda)}$ in Table \ref{tb:ne2add} is 0.993.
As the result described in Case 1 for \cite{tomizawa2004generalization}'s index, Table \ref{tb:ne2r_add} shows that the closer the value of the index is to the boundary of its range, the more frequently the improved estimates are in the outside of the range. 
In particular, in Table \ref{tb:ne2}a, where the value of the index is close to 0, and in Table \ref{tb:ne2add}, where the value is close to 1, the improved estimates are in the outside of the range of the index for about 10\% of the 10000 simulations.
\end{rem}

\begin{table}[h]
\caption{The $4 \times 5$ structures of probabilities to generate a multinomial random number}
\begin{center}
\scalebox{0.9}{
\begin{tabular}{ccccc}
\hline
\begin{tabular}{rrrrr}
\multicolumn{5}{l}{(d)} \\ \hline
$0.048$&$0.055$&$0.065$&$0.033$&$0.048$\tabularnewline
$0.032$&$0.061$&$0.048$&$0.038$&$0.068$\tabularnewline
$0.045$&$0.071$&$0.046$&$0.042$&$0.054$\tabularnewline
$0.049$&$0.052$&$0.049$&$0.033$&$0.063$\tabularnewline
\hline
\end{tabular}
&&
\begin{tabular}{rrrrr}
\multicolumn{5}{l}{(e)} \\ \hline
$0.199$&$.0005$&$.0005$&$.0005$&$0.198$\tabularnewline
$.0005$&$0.001$&$0.197$&$0.001$&$.0005$\tabularnewline
$0.001$&$0.198$&$.0005$&$.0005$&$0.001$\tabularnewline
$.0005$&$.0005$&$0.001$&$0.198$&$.0005$\tabularnewline
\hline
\end{tabular}
\end{tabular}
}
\end{center}
\label{tb:ne1add}
\end{table}%

\begin{table}[h]
\caption{The number of times that the improved estimate is in the outside of the range of \cite{tomizawa2004generalization}'s index $V^{(\lambda)}$ for Tables \ref{tb:ne1}a, \ref{tb:ne1}b, \ref{tb:ne1}c, \ref{tb:ne1add}d, and \ref{tb:ne1add}e for 10000 simulations when $\gamma = 1, 2, \ldots, 10$, where $\gamma$ is the proportion of sample size to the number of cells}
\begin{center}
\begin{tabular}{l rrrrrrrrrr} \hline
 & \multicolumn{10}{c}{$\gamma$} \\ \cline{2-11}
 & 1 & 2 & 3 & 4 & 5 & 6 & 7 & 8 & 9 & 10 \\ \hline
Table \ref{tb:ne1}a & 1034 & 3052 & 3471 & 3654 & 3545 & 3467 & 3379 & 3131 & 2962 & 2853 \\ 
Table \ref{tb:ne1}b & 224 & 245 & 68 & 22 & 4 & 3 & 0 & 0 & 0 & 0 \\ 
Table \ref{tb:ne1}c & 9 & 0 & 0 & 0 & 0 & 0 & 0 & 0 & 0 & 0 \\
Table \ref{tb:ne1add}d & 452 & 156 & 51 & 10 & 6 & 0 & 0 & 0 & 0 & 0 \\ 
Table \ref{tb:ne1add}e & 1544 & 1673 & 1657 & 1417 & 1382 & 1051 & 1367 & 617 & 585 & 739 \\ \hline
\end{tabular}
\end{center}
\label{tb:ne1r_add}
\end{table}%

\begin{table}[h]
\caption{The $4 \times 4$ structures of probabilities to generate a multinomial random number}
\begin{center}
\scalebox{1}{
\begin{tabular}{rrrr} \hline
$0.100$&$0.097$&$.0002$&$0.102$\tabularnewline
$.0002$&$0.100$&$0.098$&$.0001$\tabularnewline
$0.095$&$.0002$&$0.100$&$0.111$\tabularnewline
$.0001$&$0.096$&$.0002$&$0.100$\tabularnewline
\hline
\end{tabular}
}
\end{center}
\label{tb:ne2add}
\end{table}%

\begin{table}[h]
\caption{The number of times that the improved estimate is in the outside of the range of \cite{tomizawa1998power}'s index $\Phi^{(\lambda)}$ for Tables \ref{tb:ne2}a, \ref{tb:ne2}b, \ref{tb:ne2}c, and \ref{tb:ne2add} for 10000 simulations when $\gamma = 1, 2, \ldots, 10$, where $\gamma$ is the proportion of sample size to the number of cells}
\begin{center}
\begin{tabular}{l rrrrrrrrrr} \hline
 & \multicolumn{10}{c}{$\gamma$} \\ \cline{2-11}
 & 1 & 2 & 3 & 4 & 5 & 6 & 7 & 8 & 9 & 10 \\ \hline
Table \ref{tb:ne2}a & 369 & 1366 & 1752 & 1661 & 1374 & 1160 & 944 & 705 & 581 & 438 \\ 
Table \ref{tb:ne2}b & 31 & 50 & 32 & 8 & 0 & 1 & 0 & 0 & 0 & 0 \\ 
Table \ref{tb:ne2}c & 0 & 0 & 223 & 0 & 97 & 30 & 26 & 0 & 9 & 5 \\
Table \ref{tb:ne2add} & 0 & 0 & 1061 & 0 & 1163 & 873 & 904 & 0 & 840 & 1125 \\ \hline 
\end{tabular}
\end{center}
\label{tb:ne2r_add}
\end{table}%

%=sec:concluing=========================================================
\section{Concluding Remarks}
\label{sec:concluding}
This study solved the problem of poor estimation accuracy of the estimator of indexes with sample proportions without a sufficient number of samples, by using the Bayesian estimators of cell probabilities under the assumption of the Dirichlet prior.
In doing so, we asymptotically evaluated the MSE of the estimator of indexes with the Bayesian estimators of cell probabilities, and derived the Dirichlet parameter that minimizes the MSE.

\cite{tomizawa2007improved}, \cite{tahata2008improved}, and \cite{tahata2014refined} derived higher orders of bias and performed bias corrections, but the range of possible values for their estimators is not equal to that of their corresponding indexes, which made it difficult to interpret the values of the estimates and their confidence intervals.
In addition, since it is difficult to derive the asymptotic distribution of the improved estimator, the uncertainty quantification of indexes may not be possible.
Typically, when the true value of index is on the boundary of its range, it is difficult to quantify the uncertainty using the frequentist approach, such as the plug-in estimator with sample proportions. 
Our approach is based on the Bayesian approach, and therefore, the uncertainty can be easily quantified using the Monte Carlo method.
%Our proposed method solved this problem by using the Bayesian estimators of cell probabilities, and the evaluation of uncertainty for indexes can be simplified by using the Monte Carlo method.

Numerical experiments confirmed that our proposed estimator improves the estimation accuracy over the estimator of indexes with the sample proportions in most settings.
We also confirmed that the proposed estimator has the same estimation accuracy as the estimator of indexes even in the setting where the contingency tables contain the cells with probabilities close to zero, which is advantageous for the sample proportions.
Compared to the improved estimator with only bias correction, our proposed estimator has about the same values of the bias but slightly smaller values of the MSE.

When using the Bayesian estimators of cell probabilities, there is a question of which the Dirichlet parameter to use.
We provided one answer, which is to use the Dirichlet parameter that minimizes the MSE of the estimator of indexes in order to improve the accuracy of the estimation of indexes.
In fact, numerical experiments showed that when the uniform prior, Jeffreys prior, and the Dirichlet parameter chosen by the \cite{fienberg1973simultaneous}'s method are used, their estimation accuracies are considerably worse than that of the estimator with the sample proportions, depending on the probability structure of the contingency tables.
In contrast, as mentioned above, our proposed estimator did not become considerably worse than the estimator with the sample proportions, and our estimator maintained stable estimation accuracy in all settings.

Surprisingly, we confirmed that improving the estimator of indexes itself improves the estimation accuracy rather than using the improved estimators of cell probabilities in the estimation of indexes.
\cite{fienberg1973simultaneous} asymptotically evaluated the MSE of the Bayesian estimators of the cell probabilities and derived the Dirichlet parameter that minimizes the MSE.
Namely, they derived
\begin{equation*}
\alpha^{FH} = \argmin_{\alpha} \lim_{n\to\infty} n^2 \MSE[\hat{\bm{p}}^{(\alpha)}].
\end{equation*}
However, as shown by numerical experiments, in the estimation of indexes, the Dirichlet parameter chosen by the \cite{fienberg1973simultaneous}'s method cannot improve the estimation accuracy of indexes in many cases.

In conclusion, when inferring indexes in contingency tables, we recommend the use of our proposed estimator, which can easily evaluate the uncertainty of indexes and improves the estimation accuracy compared to the estimator of indexes with the sample proportions.

\bibliographystyle{apalike} 
\bibliography{sn-bibliography}

\begin{thebibliography}{}

\bibitem[Agresti, 2013]{agresti2013categorical}
Agresti, A. (2013).
\newblock {\em {Categorical Data Analysis}}.
\newblock John Wiley and Sons, Hoboken, New Jersey, 3rd edition.

\bibitem[Bayes, 1763]{bayes1763lii}
Bayes, T. (1763).
\newblock {An Essay Towards Solving a Problem in the Doctrine of Chances}.
\newblock {\em Philosophical transactions of the Royal Society of London},
  53:370--418.

\bibitem[Berger et~al., 2015]{berger2015overall}
Berger, J.~O., Bernardo, J.~M., and Sun, D. (2015).
\newblock {Overall Objective Priors}.
\newblock {\em Bayesian Analysis}, 10:189--221.

\bibitem[Bishop et~al., 2007]{bishop2007discrete}
Bishop, Y.~M., Fienberg, S.~E., and Holland, P.~W. (2007).
\newblock {\em {Discrete Multivariate Analysis: Theory and Practice}}.
\newblock Springer Science \& Business Media.

\bibitem[Bowker, 1948]{bowker1948test}
Bowker, A.~H. (1948).
\newblock {A Test for Symmetry in Contingency Tables}.
\newblock {\em Journal of the American Statistical Association}, 43:572--574.

\bibitem[Caussinus, 1965]{AFST_1965_4_29__77_0}
Caussinus, H. (1965).
\newblock {Contribution \`a l'analyse statistique des tableaux de
  corr\'elation}.
\newblock {\em Annales de la Facult\'e des sciences de Toulouse}, 29:77--183.

\bibitem[Cram{\'e}r, 1946]{cramir1946mathematical}
Cram{\'e}r, H. (1946).
\newblock {\em {Mathematical Methods of Statistics}}.
\newblock Princeton, N.J., Princeton Univ. Press.

\bibitem[Cressie and Read, 1984]{cressie1984multinomial}
Cressie, N. and Read, T.~R. (1984).
\newblock Multinomial goodness-of-fit tests.
\newblock {\em Journal of the Royal Statistical Society: Series B
  (Methodological)}, 46(3):440--464.

\bibitem[Fienberg and Holland, 1972]{fienberg1972choice}
Fienberg, S.~E. and Holland, P.~W. (1972).
\newblock {On the Choice of Flattening Constants for Estimating Multinomial
  Probabilities}.
\newblock {\em Journal of Multivariate Analysis}, 2:127--134.

\bibitem[Fienberg and Holland, 1973]{fienberg1973simultaneous}
Fienberg, S.~E. and Holland, P.~W. (1973).
\newblock {Simultaneous Estimation of Multinomial Cell Probabilities}.
\newblock {\em Journal of the American Statistical Association}, 68:683--691.

\bibitem[Goodman, 1979]{goodman1979multiplicative}
Goodman, L.~A. (1979).
\newblock {Multiplicative Models for Square Contingency Tables with Ordered
  Categories}.
\newblock {\em Biometrika}, 66:413--418.

\bibitem[Goodman and Kruskal, 1954]{goodman1954measures}
Goodman, L.~A. and Kruskal, W.~H. (1954).
\newblock {Measures of Association for Cross Classifications}.
\newblock {\em Journal of the American Statistical Association}, 49:732--764.

\bibitem[Jeffreys, 1946]{jeffreys1946invariant}
Jeffreys, H. (1946).
\newblock {An Invariant Form for the Prior Probability in Estimation Problems}.
\newblock {\em Proceedings of the Royal Society of London. Series A.
  Mathematical and Physical Sciences}, 186:453--461.

\bibitem[McCullagh, 1978]{10.1093/biomet/65.2.413}
McCullagh, P. (1978).
\newblock {A Class of Parametric Models for the Analysis of Square Contingency
  Tables with Ordered Categories}.
\newblock {\em Biometrika}, 65:413--418.

\bibitem[Patil and Taillie, 1982]{patil1982diversity}
Patil, G. and Taillie, C. (1982).
\newblock {Diversity as a Concept and its Measurement}.
\newblock {\em Journal of the American Statistical Association}, 77:548--561.

\bibitem[Perks, 1947]{perks1947some}
Perks, W. (1947).
\newblock {Some Observations on Inverse Probability Including a New
  Indifference Rule}.
\newblock {\em Journal of the Institute of Actuaries}, 73:285--334.

\bibitem[{R Core Team}, 2022]{r2022r}
{R Core Team} (2022).
\newblock {\em R: A Language and Environment for Statistical Computing}.
\newblock R Foundation for Statistical Computing, Vienna, Austria.

\bibitem[Stuart, 1955]{10.1093/biomet/42.3-4.412}
Stuart, A. (1955).
\newblock {A Test for Homogeneity of the Marginal Distributions in a Two-Way
  Classification}.
\newblock {\em Biometrika}, 42:412--416.

\bibitem[Tahata et~al., 2014]{tahata2014refined}
Tahata, K., Tanaka, H., and Tomizawa, S. (2014).
\newblock {Refined Estimators of Measures for Marginal Homogeneity in Square
  Contingency Tables}.
\newblock {\em International Journal of Pure and Applied Mathematics},
  90:501--513.

\bibitem[Tahata et~al., 2008]{tahata2008improved}
Tahata, K., Tomisato, R., and Tomizawa, S. (2008).
\newblock {An Improved Approximate Unbiased Estimator of Log-Odds Ratio for
  2×2 Contingency Tables}.
\newblock {\em Advances and Applications in Statistics}, 9:1--12.

\bibitem[Tahata and Tomizawa, 2014]{tahata2014symmetry}
Tahata, K. and Tomizawa, S. (2014).
\newblock {Symmetry and Asymmetry Models and Decompositions of Models for
  Contingency Tables}.
\newblock {\em SUT Journal of Mathematics}, 50:131--165.

\bibitem[Tomizawa and Makii, 2001]{tomizawa2001generalized}
Tomizawa, S. and Makii, K. (2001).
\newblock {Generalized Measures of Departure from Marginal Homogeneity for
  Contingency Tables with Nominal Categories}.
\newblock {\em Journal of Statistical Research}, 35:1--24.

\bibitem[Tomizawa et~al., 2007]{tomizawa2007improved}
Tomizawa, S., Miyamato, N., and Ohba, N. (2007).
\newblock {Improved Approximate Unbiased Estimators of Measure of Asymmetry for
  Square Contingency Tables}.
\newblock {\em Advances and Applications in Statistics}, 7:47--63.

\bibitem[Tomizawa et~al., 2001]{tomizawa2001theory}
Tomizawa, S., Miyamoto, N., and Hatanaka, Y. (2001).
\newblock {Measure of Asymmetry for Square Contingency Tables Having Ordered
  Categories}.
\newblock {\em Australian and New Zealand Journal of Statistics}, 43:335--349.

\bibitem[Tomizawa et~al., 2004]{tomizawa2004generalization}
Tomizawa, S., Miyamoto, N., and Houya, H. (2004).
\newblock {Generalization of Cramer's Coefficient of Association for
  Contingency Tables}.
\newblock {\em South African Statistical Journal}, 38:1--24.

\bibitem[Tomizawa et~al., 2005]{tomizawa2005power}
Tomizawa, S., Miyamoto, N., and Yamane, S. (2005).
\newblock {Power-Divergence-Type Measure of Departure from Diagonals-Parameter
  Symmetry for Square Contingency Tables with Ordered Categories}.
\newblock {\em Statistics}, 39:107--115.

\bibitem[Tomizawa et~al., 1997]{tomizawa1997generalized}
Tomizawa, S., Seo, T., and Ebi, M. (1997).
\newblock {Generalized Proportional Reduction in Variation Measure for Two-Way
  Contingency Tables}.
\newblock {\em Behaviormetrika}, 24:193--201.

\bibitem[Tomizawa et~al., 1998]{tomizawa1998power}
Tomizawa, S., Seo, T., and Yamamoto, H. (1998).
\newblock {Power-Divergence-Type Measure of Departure from Symmetry for Square
  Contingency Tables that Have Nominal Categories}.
\newblock {\em Journal of Applied Statistics}, 25:387--398.

\bibitem[Tuyl, 2019]{tuyl2018method}
Tuyl, F. (2019).
\newblock {A Method to Handle Zero Counts in the Multinomial Model}.
\newblock {\em The American Statistician}, 73:151--158.

\bibitem[Yule, 1900]{yule1900vii}
Yule, G.~U. (1900).
\newblock {On the Association of Attributes in Statistics}.
\newblock {\em Philosophical Transactions of the Royal Society of London.
  Series A, Containing Papers of a Mathematical or Physical Character},
  194:257--319.

\bibitem[Yule, 1912]{yule1912methods}
Yule, G.~U. (1912).
\newblock {On the Methods of Measuring Association Between Two Attributes}.
\newblock {\em Journal of the Royal Statistical Society}, 75:579--652.

\end{thebibliography}
\end{document}